\documentclass[aps,prd,amsfonts,preprint,eqsecnum]{revtex4-1}
\usepackage{bm}
\usepackage{mathrsfs}
\usepackage{natbib}
\usepackage{graphicx}
\usepackage{amsmath,amssymb,amstext}
\usepackage{epstopdf}
\usepackage{subfigure}
\usepackage{tensor}

\newcommand{\fv}[1]{\bm{#1}}
\newcommand{\gv}[1]{\nabla_{\bm{#1}}}

\newcommand{\pphi}[0]{{\partial_{\phi}}}
\newcommand{\plambda}[0]{{\partial_{\lambda}}}
\newcommand{\pert}[1]{\Delta^1_\varphi\left[#1\right]}

\begin{document}
\title{Gravitational wave detector response in terms of spacetime Riemann curvature}
\date{\today}

\author{Michael J. Koop}
\affiliation{Department of Physics, 104 Davey Laboratory, The Pennsylvania State University, University Park, PA  16802-6300, USA}

\author{Lee Samuel Finn}
\email{lsfinn@psu.edu}
\affiliation{Department of Physics and Department of Astronomy \& Astrophysics, The Pennsylvania State University, University Park, PA  16802-6300, USA}

%\input{abstract.tex}
%!TEX root = ./ms.tex
% $Id: abstract.tex 6952 2013-10-01 15:01:22Z mkoop $
\begin{abstract} 

Gravitational wave detectors are typically described as responding to gravitational wave metric perturbations, which are gauge-dependent and --- correspondingly --- unphysical quantities. This is particularly true for ground-based interferometric detectors, like LIGO, space-based detectors, like LISA and its derivatives, spacecraft doppler tracking detectors, and pulsar timing arrays detectors. 
The description of gravitational waves, and a gravitational wave detector's response, to the unphysical metric perturbation has lead to a proliferation of false analogies and descriptions regarding how these detectors function, and true misunderstandings of the physical character of gravitational waves. 

Here we provide a fully physical and gauge invariant description of the response of a wide class of gravitational wave detectors in terms of the Riemann curvature, the physical quantity that describes gravitational phenomena in general relativity. 
In the limit of high frequency gravitational waves, the Riemann curvature separates into two independent gauge invariant quantities: a ``background'' curvature contribution and a ``wave'' curvature contribution. In this limit the gravitational wave contribution to the detector response reduces to an integral of the gravitational wave contribution of the curvature along the unperturbed photon path between components of the detector. The description presented here provides an unambiguous physical description of what a gravitational wave detector measures and how it operates, a simple means of computing corrections to a detectors response owing to general detector motion, a straightforward way of connecting the results of numerical relativity simulations to gravitational wave detection, and a basis for a general and fully relativistic pulsar timing formula. 
\end{abstract}

% 04.80.Nn	Gravitational wave detectors and experiments (see also 95.55.Ym Gravitational radiation detectors; mass spectrometers; and other instrumentation and techniques)
% 95.55.Ym	Gravitational radiation detectors; mass spectrometers; and other instrumentation and techniques (see also 04.80.Nn Gravitational wave detectors and experiments in—General relativity and gravitation)
% 95.30.Sf	Relativity and gravitation (see also section 04 General relativity and gravitation; 98.80.Jk Mathematical and relativistic aspects of cosmology)
% 04.30.Nk	Wave propagation and interactions

\maketitle

%\input{intro.tex}
%!TEX root = ./ms.tex
% $Id: intro.tex 6958 2013-10-09 17:10:35Z lsfinn $
\section{Introduction}\label{sec:intro}
It has long been understood that, in general relativity gravitational phenomena are the physical manifestation of spacetime curvature. Nevertheless, gravitational wave detectors, which make physical measurements, are typically described as 
responding to spacetime metric perturbations, which are coordinate gauge dependent and --- correspondingly --- unphysical quantities. Just as early attempts to understand gravitational waves in terms of metric perturbations led to confusion regarding whether such waves existed or how they might be generated, so attempts to describe how gravitational wave detectors respond to metric perturbations lead to wooly statements and, sometimes, outright misconceptions \cite{faraoni:1992gu,saulson:1997:ilw,garfinkle:2006:gia,faraoni:2007:cma,finn:2009:roi,kennefick:2007:tas}. By way of contrast, gravitational waves described as spacetime curvature perturbations are, in a well-defined sense \cite{isaacson:1968:gri}, physically unambiguous quantities; correspondingly, describing the gravitational wave detector response in terms of the detector's interaction with spacetime curvature may be expected to be, if nothing else, conceptually more satisfying and physically more revealing. Here we derive and describe the response of a wide class of gravitational wave antennas --- including pulsar timing arrays, spacecraft Doppler tracking, and both ground- and space-based laser interferomteric detectors --- in a way that relies solely upon physical measurements and the physical properties of spacetime as described by the Riemann curvature tensor. 
The resulting expression of the detector response, and each term that comprises it, is separately gauge invariant and has a clear physical interpretation.
We show that when the gravitational waves can be described as a gauge independent curvature perturbation of a background spacetime the wave contribution to the response is wholly embodied in an integration of a projection of Riemann curvature tensor perturbation along specific null geodesics of the unperturbed spacetime. Our principle result is directly applicable to gravitational wave detection via pulsar timing or spacecraft Doppler tracking; is the building-block upon which a physically and pedagogically satisfying description of the response of interferometeric antennas may be based; can form the basis for a simplified and general derivation of a fully relativistic pulsar timing formula; and may be used to simplify the use of numerial relativity simulations to aid in the analysis or interpretation of gravitational wave detector observations. 

\citet{weber:1960:dag} described the operation of the first practical gravitational wave detector in terms of Riemann curvature induced excitations of the oscillation modes of a high quality-factor metal bar. Throughout the 1960's a variety of mechanical gravitational wave antenna configurations were proposed (a qualitative summary of many of these configurations is given in \cite[pg.~1013]{misner:1973:g}) and their responses were generally described in an approximate way through a coupling to the Riemann curvature as well. By the early 1970s, however, a description of gravitational waves in terms of metric perturbations had taken hold. This was the case from the start for interferometric gravitational wave antennas \cite{forward:1978:wlg,weiss:1972:ecb} and their kin: spacecraft doppler tracking \cite{estabrook:1975:rod} and pulsar timing \cite{sazhin:1978:ofd,detweiler:1979:ptm}. This trend of describing gravitational wave antenna response to metric perturbations has generally persisted throughout the literature to this day \cite{2007PhRvL..98k1102D,2008PhRvD..78d2003D,2008PhRvD..78l2002D,Graham:2013ek}.

As is well known, however, metric perturbations depend on coordinate gauge choices. While the freedom to choose a gauge may be exploited to simplify some computations, it is perilous to give physical meaning to partial or intermediate results of these calculations. Nevertheless, throughout the literature --- research and pedagogical --- one finds numerous (and sometimes conflicting) descriptions of how a gravitational wave physically produces a signal in a detector based on an interpretation of the terms in calculations that depend on a choice of coordinate gague. For example, it is routinely claimed (though we give only single, recent examples from the literature) that 
gravitational waves physically move freely-falling test masses in an ideal interferometric detector \cite[pg.~21]{Saulson:2006wg}; 
gravitational waves ``push'' the mirrors of an ideal interferometric detector apart and together \cite[Lecture 1 page 1]{thorne:2002:gwa}; 
spacetime curvature influences light differently than it does mirror separations in an interferometer \cite[Lecture 1 page 2]{thorne:2002:gwa}; 
gravitational waves alternately redshift and blueshift the light in an interferometer detector \cite{1997AmJPh..65..501S};
it is the direct affect of gravitational waves on the Earth and a distant pulsar that we measure when we detect when we observe gravitational waves via pulsar timing \cite{Cordes:2012cq}; 
and it is the affect of gravitational waves on the trajectory of the light passing between a pulsar and Earth that leads to their detection \cite{anholm:2009:osf}.    
In fact, as is apparent by appropriate application of the Equivalence Principle, each of these statements is incorrect; that they are made at all is the result of ascribing physical significance to gauge-dependent quantities. 

Other authors have explored alternative derivations or expressions of the gravitational wave detector response in terms of the Riemann curvature \cite{garfinkle:2006:gia,faraoni:2007:cma,schutz:1990:fci,creighton:2009:pta}; however, these discussions either start from the metric perturbation in a preferred gauge or apply, in an approximate way, the geodesic deviation equation to describe the deviation vector between non-geodesics. Additionally, these descriptions (as well as most formulations coupling to the metric perturbation) generally make the crucial assumption that the background spacetime is Minkowski or that the different detector components (beamsplitter and end mirrors for interferometers or the Earth and a pulsar for pulsar timing) are at some coordinate rest in a preferred gauge. The description of the response we present here makes no assumptions regarding the geometry of the background spacetime and involves, from beginning to end, only physical measurements and gauge-independent quantities: i.e., it is valid in an arbitrary background spacetime and never requires or invokes any special gauge or gauge-dependent quantities. 

In Section \ref{sec:doppler} we provide a general, geometrically motivated derivation for the observed phase evolution of a remote clock. In Section \ref{sec:gwaves} we review how and when spacetime curvature may be physically and unambiguously separated into background and gravitational wave perturbation contributions, each of which is separately gauge independent, and show that, when such a distrinction is possible, the gravitational wave contribution to the observed clock phase is entirely due to the gravitational wave contribution to the curvature. We discuss the meaning and application of our results in Section \ref{sec:discussion}. We end in Section \ref{sec:conclusions} with some brief conclusions and directions for future study. 

%\input{doppler.tex}
%!TEX root = ./ms.tex
% $Id: doppler.tex 6952 2013-10-01 15:01:22Z mkoop $

\section{The observed phase evolution of a remote clock}\label{sec:doppler}

Gravitational wave detection via pulsar timing \cite{sazhin:1978:ofd,detweiler:1979:ptm,backer:1986:pta}, spacecraft Doppler tracking \cite{estabrook:1975:rod,hellings:1981:sgd,armstrong:2006:lgw}, and ground- or space-based laser interferometry \cite{lrr-2010-1,lrr-2011-5,estabrook:2000:tao,armstrong:2001:soa} all involve measuring the phase of a remote clock via an electromagnetic signal that propagates through spacetime along (piecewise) null geodesics. The location of the clock, the receiver/observer, and the path taken by the electromagnetic record of the phase as it is transferred from the clock to the receiver/observer differ; however, the measured record of the clock phase following its propagation along the null trajectory is the defining characteristic of these detectors. In this section we develop a fully relativistic expression for evolution of that record of the clock's phase in terms of the clock's intrinsic phase evolution, the clock and receiver/observer spacetime trajectories, the null geodesic trajectories connecting the clock and observer, and the spacetime curvature along those geodesics. In section \ref{sec:gwaves} we separate-out from this expression just that part of the response due to gravitational waves. 

\subsection{Observed clock phase}\label{subsec:geodev}
\begin{figure}
\includegraphics[width=0.5\textwidth]{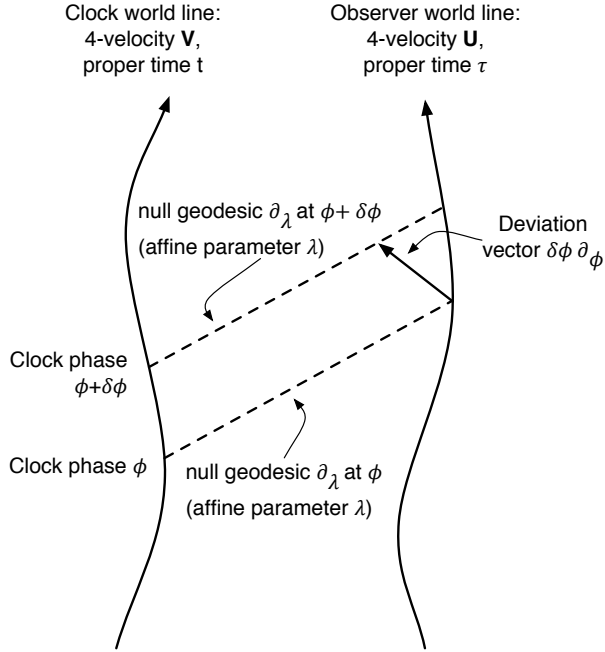}
\caption{Space-time diagram illustrating transmission of clock phase from clock to receiver/observer.}\label{fig:geoDev}
\end{figure}

Referring to Figure \ref{fig:geoDev}, identify the world lines of the clock and the receiver/observer. Denote the clock's 4-velocity $\bm{V}$ and the observer's 4-velocity $\bm{U}$. Focus attention on the future-directed null geodesic congruence that connects the clock's world line (in the past) and the receiver/observer's world line (in the future). Identify each null geodesic in the congruence by the clock's (monatonically increasing) phase $\phi$ on the clock's world line where it intersects that null geodesic. Let $\lambda$ be an affine parameter distance along each null-geodesic measured from the clock's world line. In this way $\phi$ and $\lambda$ are coordinate functions on the world sheet described by the null geodesic congruence. 

Next note that, on the receiver/observer's world line, 
\begin{align}\label{eq:U}
\fv{U} &= \frac{d\phi}{d\tau}\pphi + \frac{d\lambda}{d\tau}\plambda\,,
\end{align}
where $\plambda$ and $\pphi$ are the coordinate vectors corresponding to the coordinate functions $\lambda$ and $\phi$. The \emph{observed} rate of phase is thus
\begin{align}\label{eq:dPhi:dTau}
\frac{d\phi}{d\tau} &= \frac{\fv{U}\cdot\plambda}{\pphi\cdot\plambda}.
\end{align}

Evaluating the right-hand side of Equation \ref{eq:dPhi:dTau}
requires knowing how $\pphi$ varies along the geodesics described by $\plambda$. Noting that the coordinate vector field $\partial_{\lambda}$ is a null geodesic congruence whose deviation vectors are  $\pphi$ we can solve the equation of geodesic deviation to find $\pphi$ along $\plambda$. 

For our purposes it is convenient to write the geodesic deviation equation as a set of first-order equations
\begin{subequations} \label{eq:K}
\begin{align}
\gv{\plambda}\pphi &= \fv{K}\\
\gv{\plambda}\fv{K} &= -\fv{R}(\ldots,\plambda,\pphi,\plambda).
\end{align}
\end{subequations}
To complete the specification of the problem requires boundary conditions on $\pphi$ and $\bm{K}$. We set the boundary conditions on $\pphi$ on the clock's world line, where 
\begin{align}
\left.\pphi\right|_{\lambda=0} &=\left(\frac{d\phi}{dt}\right)^{-1}\fv{V},
\end{align}
$\fv{V}=\partial_t$ is the clock's 4-velocity, and $d\phi/dt$ is the clock's instantaneous frequency. The boundary condition on $\bm{K}$ can be set in terms of the receiver/observer's trajectory and the observed direction to the clock. To do so, note that, since $\phi$ and $\lambda$ are coordinate functions, 
\begin{align}
\bm{K} &= \gv{\plambda}\pphi = \gv{\pphi}\plambda\,.\label{eq:kFlip}
\end{align}
Next note that, on the receiver/observer's world line, 
\begin{align}\label{eq:plambda}
\plambda&= -\left(\fv{U}-\fv{n}\right)\fv{U}\cdot\plambda\,.
\end{align}
where $\fv{n}$ is the apparent direction to the clock as measured in the receiver/observer's proper reference frame. Combining Equations \ref{eq:kFlip}, \ref{eq:U} and \ref{eq:plambda}, find
\begin{align}
\left.\bm{K}\right|_{\lambda=\lambda_R} &= -\left(\frac{d\tau}{d\phi}\right)\gv{U}\left[\left(\fv{U}-\fv{n}\right){\fv{U}\cdot\plambda}\right]\,.
\end{align}
Lastly, take advantage of the freedom to choose the affine parameter scale to set
\begin{align}\label{eq:lambdaNormalization}
\fv{U}\cdot\plambda &= -1,
\end{align}
thereby obtaining the boundary condition
\begin{align}
\left.\bm{K}\right|_{\lambda=\lambda_R} &= 
\left(\frac{d\phi}{d\tau}\right)^{-1}
\left(\fv{a}_R-\dot{\fv{\theta}}\right)
\end{align}
where $\fv{a}_R$ is the receiver/observer's 4-acceleration and $\dot{\fv{\theta}}=\gv{U}\fv{n}$ is the angular velocity (proper motion) of the clock as measured by the receiver/observer. 

To recapitulate, the clock phase record as observed at the receiver evolves as
\begin{align}
\frac{d\phi}{d\tau} &= \frac{\fv{U}\cdot\plambda}{\pphi\cdot\plambda},
\end{align}
where $\fv{U}$ is the receiver's 4-velocity, $\plambda$ is the tangent to the null geodesic along which the clock phase record is propagated and $\pphi$ is the deviation associated with the null geodesic field $\plambda$. The deviation vectors are computed via
\begin{subequations}
the equation of geodesic deviation
\begin{align}
\gv{\plambda}\pphi &= \fv{K}\\
\gv{\plambda}\fv{K} &= -\bm{R}(\cdots,\plambda,\pphi,\plambda)\,,
\end{align}
with boundary conditions and null-geodesic affine parameter normalization
\begin{align}
\left.\pphi\right|_{\lambda=0} &=\left(\frac{d\phi}{dt}\right)^{-1}\fv{V}\,,
\label{eq:bcPhi}\\
\left.\bm{K}\right|_{\lambda=\lambda_R} &= 
\left(\frac{d\phi}{d\tau}\right)^{-1}
\left(\fv{a}_R-\dot{\fv{\theta}}\right)\,
\label{eq:bcLambda}\\
\fv{U}\cdot\plambda &= -1\,.
\label{eq:lambdaScale}
\end{align}
\end{subequations}

\subsection{Observed clock Doppler}\label{sec:diffEq}
A more physically revealing description of the observed clock phase may be found by focusing on the clock phase's observed Doppler, or $d^2\phi/d\tau^2$. Our objective in this subsection is to find the observed Doppler  and show that it can be expressed in terms of the receiver/observer's acceleration, the clock's intrinsic frequency and frequency derivative ($d\phi/dt$ and $d^2\phi/dt^2$), the clock's apparent transverse velocity and --- most importantly for the gravitational wave signal --- the curvature along the null geodesics connecting the clock and receiver/observer. 

Begin by adopting the affine parameter normalization in Equation \ref{eq:lambdaScale} and writing the observed clock Doppler as 
\begin{subequations}\label{eq:doppler}
\begin{align}
\frac{d^2\phi}{d\tau^2} &= 
\gv{U}\left(\frac{\plambda\cdot\fv{U}}{\plambda\cdot\pphi}\right)\\
&= 
\left(\plambda\cdot\pphi\right)^{-2}\left[\frac{d\phi}{d\tau}\pphi + \frac{d\lambda}{d\tau}\plambda\right]\cdot\nabla\left(\plambda\cdot\pphi\right)\\
&= \left(\frac{d\phi}{d\tau}\right)^3
\left[\gv{\pphi}\left(\plambda\cdot\pphi\right)\right]_{\lambda=\lambda_R} 
+ \left[\left(\frac{d\phi}{d\tau}\right)^2
  \left(\frac{d\lambda}{d\tau}\right)
  \left(\plambda\cdot\fv{K}\right)\right]_{\lambda=\lambda_R}\\
&= \left(\frac{d\phi}{d\tau}\right)^3
\left[
\gv{\pphi}
\left(\plambda\cdot\pphi\right)
\right]_{\lambda=\lambda_R}\,.
\label{eq:dphi}
\end{align}
\end{subequations}
In arriving at Equation \ref{eq:dphi} we have taken advantage of the affine parameter normalization (Eq.~\ref{eq:lambdaScale}), the resolution of $\fv{U}$ in terms of the ($\pphi$, $\plambda$) basis (Eq.~\ref{eq:U}), and the orthogonality of $\plambda$ and $\fv{K}$:
\begin{align}\label{eq:KOrthogonality}
\plambda\cdot\fv{K} &= \plambda\cdot\left(\gv{\pphi}\plambda\right) = 0
\end{align}

Next note that, since $\fv{K}$ and $\plambda$ are orthogonal (Eq.~\ref{eq:KOrthogonality}), 
\begin{subequations}
\begin{align}
\gv{\plambda}\left(\plambda\cdot\pphi\right) &= \plambda\cdot\left(\gv{\plambda}\pphi\right) = \plambda\cdot\fv{K} = 0.
\end{align}
Also
\begin{align}\label{eq:pphiConst}
\gv{\plambda}\gv{\phi}\left(\plambda\cdot\pphi\right) &= \gv{\phi}\gv{\plambda}\left(\plambda\cdot\pphi\right) = 0\,;
\end{align}
\end{subequations}
i.e., both $\plambda\cdot\pphi$ and $\gv{\pphi}\left(\plambda\cdot\pphi\right)$ are constant along the null geodesics of the congruence. Thus
\begin{subequations}\label{eq:retarded}
\begin{align}
\left.\gv{\pphi}\left(\plambda\cdot\pphi\right)\right|_{\lambda=\lambda_R} &= 
\left.\gv{\pphi}\left(\plambda\cdot\pphi\right)\right|_{\lambda=0} \\
\left.\plambda\cdot\pphi\right|_{\lambda=\lambda_R} &= \left.\plambda\cdot\pphi\right|_{\lambda=0}
\end{align}
\end{subequations}
with the understanding that the right- and left-hand sides of Equation \ref{eq:retarded} are evaluated on the same null geodesic. 

Taking advantage of Equation~\ref{eq:retarded} expand the bracketed term in Equation \ref{eq:dphi}: 
\begin{subequations}
\begin{align}
\label{eq:bracketed:a}
\gv{\pphi}(\plambda\cdot\pphi)_{\lambda=\lambda_R} &= 
\left[
  \pphi\cdot\gv{\pphi}\plambda
  + \plambda\cdot\gv{\pphi}\pphi
\right]_{\lambda=0}\\
\label{eq:bracketed:b}
&= 
\left(\pphi\cdot\fv{K}\right)_{\lambda=0} 
+ \frac{
  \plambda\cdot\fv{a}_C
  }{
  \left(d\phi/dt\right)^2
  }
+ \frac{
    \plambda\cdot\pphi
    }{
    d\phi/dt
    }
  \gv{V}\left(\frac{d\phi}{dt}\right)^{-1}\\
\label{eq:bracketed:c}
&= 
\frac{\plambda\cdot\fv{a}_C}{\left(d\phi/dt\right)^2} 
+ \left(\frac{d\phi}{d\tau}\right)^{-1}\left(\frac{d\phi}{dt}\right)^{-2}
  \frac{d^2\phi}{dt^2}
  \nonumber\\
& \qquad\qquad{} + \left.\left(\pphi\cdot\fv{K}\right)\right|_{\lambda=\lambda_R}
- \int_0^{\lambda_{R}}\left[\fv{K}^2-\fv{R}(\pphi,\plambda,\pphi,\plambda)\right]d\lambda\,.
\end{align}
\end{subequations}
In proceeding from Equation \ref{eq:bracketed:a} to \ref{eq:bracketed:b} use the boundary condition on $\pphi$ (Eq.~\ref{eq:bcPhi}); in passing on to Equation \ref{eq:bracketed:c} use
\begin{align}
\left.\pphi\cdot\fv{K}\right|_{\lambda=\lambda_R}
- \left.\pphi\cdot\fv{K}\right|_{\lambda=0} &= 
\int_0^{\lambda_R}
\gv{\plambda} \left( \pphi \cdot \fv{K} \right) d\lambda
\nonumber\\
&= \int_0^{\lambda_R}
\left[\fv{K}^2 - \fv{R}(\pphi,\plambda,\pphi,\plambda)\right]d\lambda\,,
\label{eq:int}
\end{align}
where the integration path is along a constant $\phi$ null geodesic.

The boundary term $\left(\pphi\cdot\bm{K}\right)_{\lambda=\lambda_R}$ may be re-expressed in terms of the receiver/observer's 4-acceleration:
\begin{subequations}\label{eq:bterm}
\begin{align}
\label{eq:bterm:a}
\left(\pphi\cdot\bm{K}\right)_{\lambda=\lambda_R} 
&= \left(\pphi\cdot\gv{\pphi}\plambda\right)_{\lambda=\lambda_R}\\
\label{eq:bterm:b}
&= 
\left(\frac{d\phi}{d\tau}\right)^{-2}\bm{U}\cdot\gv{U}\plambda\\
\label{eq:bterm:c}
&= 
-\left(\frac{d\phi}{d\tau}\right)^{-2}
\plambda\cdot\gv{U}\fv{U}
=
-\left(\frac{d\phi}{d\tau}\right)^{-2}
\plambda\cdot\fv{a}_{R}
\end{align}
\end{subequations}
In passing from Equation \ref{eq:bterm:a} to \ref{eq:bterm:c} take advantage of the expression for $\fv{U}$ resolved on the $(\pphi,\plambda)$ coordinate basis (Eq.~\ref{eq:U}) and the choice of affine parameter normalization (Eq.~\ref{eq:lambdaScale}). 

Bringing together Equations \ref{eq:dphi}, \ref{eq:bracketed:c} and \ref{eq:bterm:c}, find
\begin{align}\label{eq:RiemannResponse}
\frac{d^2\phi}{d\tau^2} &= 
\left(\frac{d\phi/d\tau}{d\phi/dt}\right)^2
\frac{d^2\phi}{dt^2}
-
\frac{d\phi/d\tau}{\plambda\cdot\fv{U}}\left\{
\left(\frac{d\phi/d\tau}{d\phi/dt}\right)^2
\left.\left(\plambda\cdot\fv{a}_C\right)\right|_{\lambda=0}
- \left.\left(\plambda\cdot\fv{a}_R\right)\right|_{\lambda=\lambda_R}
\vphantom{
\left(\frac{d\phi/d\tau}{d\phi/dt}\right)^2
\left.\left(\plambda\cdot\fv{a}_C\right)\right|_{\lambda=0}
- \left.\left(\plambda\cdot\fv{a}_R\right)\right|_{\lambda=\lambda_R}
-\left(\frac{d\phi}{/d\tau}\right)^2
\int_0^{\lambda_{R}}
\left[\fv{K}^2-\fv{R}(\pphi,\plambda,\pphi,\plambda)\right]d\lambda
}
\right.
\nonumber\\
&\qquad\qquad
\left.
\vphantom{
\left(\frac{d\phi/d\tau}{d\phi/dt}\right)^2
\left.\left(\plambda\cdot\fv{a}_C\right)\right|_{\lambda=0}
- \left.\left(\plambda\cdot\fv{a}_R\right)\right|_{\lambda=\lambda_R}
-\left(\frac{d\phi}{/d\tau}\right)^2
\int_0^{\lambda_{R}}
\left[\fv{K}^2-\fv{R}(\pphi,\plambda,\pphi,\plambda)\right]d\lambda
}
-\left(\frac{d\phi}{d\tau}\right)^2
\int_0^{\lambda_{R}}
\left[\fv{K}^2-\fv{R}(\pphi,\plambda,\pphi,\plambda)\right]d\lambda
\right\}\,
\end{align}
where we have relaxed the assumption that $\plambda\cdot\fv{U}=-1$. In Equation \ref{eq:RiemannResponse}  all quantities are understood to be evaluated on the same $\phi$-constant geodesic of the congruence.

Equation \ref{eq:RiemannResponse} relates the Doppler in the clock's phase record to the clock's intrinsic properties ($d\phi/dt$, $d^2\phi/dt^2$) and trajectory ($\fv{a}_C$), the  trajectory of the sensing apparatus ($\fv{U}\cdot\plambda$ and $\fv{a}_R$), the relative motion between the clock and receiver/observer expressed in the changing null geodesic traveled by the record as it propagates from the clock to the measuring apparatus [$\int \fv{K}^2 d\lambda$] and the spacetime properties along that null geodesic [$\int\fv{R}(\pphi,\plambda,\pphi,\plambda)d\lambda$]. 

In section \ref{sec:gwaves} we identify when the integral of $\fv{R}(\pphi,\plambda,\pphi,\plambda)$ appearing in Equation \ref{eq:RiemannResponse} can be separated, in a gauge invariant fashion, into a contribution that is associated with the gravitational waves and another that is associated with the background spacetime curvature.

%\input{gravrad.tex}
%!TEX root = ./ms.tex
% $Id: gravrad.tex 6952 2013-10-01 15:01:22Z mkoop $
\section{Antenna response to gravitational waves} \label{sec:gwaves}

\subsection{Introduction}

Equation \ref{eq:RiemannResponse} describes how the clock's observed Doppler, determined from the record propagated from clock to measuring apparatus, depends on the clock's intrinsic frequency and frequency evolution, the 4-velocity and 4-acceleration of the clock and measuring apparatus, and the Riemann curvature along the null geodesics connecting the clock and measuring apparatus. Included are gravitational wave effects. In this section we show that the gravitational wave effects are, when the radiation wavelength is much shorter than the scale of the background curvature, wholly due to the perturbation to the Riemann integrated along the \emph{unperturbed} null geodesic trajectory linking the clock and receiver/observer. 

The first difficulty to overcome is the identification of gravitational radiation as a distinct spacetime phenomena. In general relativity all truly gravitational phenomena - including gravitational radiation - are described by spacetime's Riemann curvature tensor. Owing to the non-linear nature of the Einstein equations a complete distinction between stationary and dynamical gravitational phenomena is, in general, impossible. Nevertheless, Isaacson \cite{isaacson:1968:gri,isaacson:1968:gri:1} showed that gravitational waves emerge naturally as a distinct physical phenomena when a multi-scale perturbation expansion of the Riemann curvature is possible. To this end, in subsection \ref{sec:mpt} we briefly review the necessary elements of perturbation theory: i.e., how the choice of a bijective map (gauge) allows us to give meaning to background and perturbation, and how the perturbation changes when we change the gauge. In subsection \ref{sec:gwPert} we show that a short wavelength perturbation in the Riemann curvature is a tensor (i.e., it is gague invariant and, correspondingly, physically unambiguous) that satisfies the wave equation on the background spacetime: i.e., it \emph{is} a gravitational wave. 

Having identified gravitational waves as a gauge-invariant, physical phenomena, we next ask how the introduction of a gravitational wave perturbation affects $d^2\phi/d\tau^2$ as given in Equation \ref{eq:RiemannResponse}. In addition to its contribution to the Riemann curvature, the gravitational wave perturbation also affects the null-geodesics along which the record of the clock's phase propagates: i.e., the gravitational wave perturbation affects, in principle, every term that appears on the righthand side of Equation \ref{eq:RiemannResponse}.  In subsection \ref{sec:gwResp} we find our principal result: the dominant contribution to the observed evolution of the clock phase $d^2\phi/d\tau^2$ owing to a  gravitational wave perturbation is entirely due to the gravitational wave perturbation to the Riemann curvature integrated along the unperturbed null geodesic trajectory linking the clock and the receiver/observer.

\subsection{Gravitational waves as short wavelength perturbations of the Riemann curvature}\label{sec:shortWave}

\subsubsection{Metric perturbation theory: Review}\label{sec:mpt}

In general relativity, spacetime is a manifold with a metric. When we perturb spacetime we perturb the manifold and metric. To compare tensors on the perturbed and unperturbed manifolds --- i.e., to identify and study the perturbation --- requires a one-one and onto map between the two manifolds. Following \cite{brizuela:2009:xca} consider a one-dimensional family of spacetime manifolds and metrics, $(\widetilde{\mathcal{M}},\widetilde{\bm{g}})$, smoothly depending on the dimensionless parameter $\epsilon$. Refer to the $\epsilon=0$ manifold and metric as the background and write them without the tilde: i.e., write $\mathcal{M}$ for $\widetilde{\mathcal{M}}(0)$ and $\bm{g}$ for $\widetilde{\bm{g}}(0)$. To compare tensor fields on $\widetilde{\mathcal{M}}(\epsilon)$ with their background counterparts introduce a pull-back $\varphi^*_{\epsilon}:\widetilde{\mathcal{M}}(\epsilon)\rightarrow\mathcal{M}$. The pull-back of any tensor $\widetilde{\bm{T}}(\epsilon)$ can be written as a Taylor expansion in $\epsilon$:
\begin{align}
\varphi^*_\epsilon\widetilde{\bm{T}}(\epsilon) &= \bm{T} + 
\sum_{k=1}\frac{\epsilon^k}{k!}\Delta^{k}_{\varphi}[\bm{T}]
\end{align}
where
\begin{align}\label{eq:defPert}
\Delta_{\varphi}^k\left[\bm{T}\right] &= \left.\frac{d^k\varphi^*_{\epsilon}\widetilde{\bm{T}}}{d\epsilon^k}\right|_{\epsilon=0}.
\end{align}
We identify $\Delta^k_{\varphi}[\bm{T}]$ as the order $\epsilon^k$ perturbation of $\bm{T}$. 

The pull-back of $\widetilde{T}(\epsilon)$ on $\widetilde{\mathcal{M}}(\epsilon)$ is a field on $\mathcal{M}$. \emph{In particular, covariant derivatives of $\varphi^*_\epsilon\widetilde{\bm{T}}$, or of $\bm{T}$ and its perturbations $\Delta_\varphi^k[\bm{T}]$, are always taken with respect to background metric $g\indices{_{ab}}$.}

For our application we are particularly interested in the perturbation expansion of the metric and the Riemann tensor (and its contractions), which we write
\begin{subequations}
\begin{align}
\varphi_\epsilon^*\widetilde{\fv{g}} &= \fv{g} + \epsilon\fv{h}^{(1)} + \mathcal{O}(\epsilon^2)\\
\varphi_\epsilon^*\widetilde{\fv{R}} &= \fv{R} + \epsilon\fv{R}^{(1)} + \mathcal{O}(\epsilon^2),
\end{align}
where we have introduced the additional shorthand
\begin{align}
\fv{h}^{(1)} &= \Delta^1_{\varphi}[\fv{g}]\\
\fv{R}^{(1)} &= \Delta^1_{\varphi}[\fv{R}].
\end{align}
\end{subequations}

Physical quantities don't depend on coordinate choices; correspondingly, $\bm{T}$ is physical only if it has a measurable effect that is independent of the choice of coordinates used to measure it. Consider a one-dimensional family of coordinate transformation $\psi_t$ generated by the vector field $\xi^a$: i.e., for coordinate function $x^\mu(\mathcal{P})$
\begin{subequations}
\begin{align}
\psi_t^*x^\mu(\mathcal{P}) &= x^\mu(\psi_t(\mathcal{P}))
\end{align}
with 
\begin{align}
\psi_0^*x^\mu &= x^\mu\\
\left.\frac{d\left(\psi^*_tx^{\mu}\right)}{dt}\right|_{t=0} &= \xi^\mu.
\end{align}
\end{subequations}
\citet{bruni:1997:pos} showed that the transformation of $\fv{T}$ under $\psi^*_t$ is described by the expansion
\begin{align}
\psi^*_t \fv{T} &= \sum_{k=0}^\infty \frac{t^k}{k!}
\mathcal{L}_{\xi}^k \fv{T},
\end{align}
where $\mathcal{L}_{\fv{\xi}}$ denotes the Lie derivative with respect to $\fv{\xi}$:
\begin{align}
\mathcal{L}_{\xi}\bm{T} &= \lim_{t\rightarrow0}\frac{1}{t}\left(\psi_{t}^*\bm{T}-\bm{T}\right)
\end{align}
Correspondingly, we conclude that $\bm{T}$ is physical only if $\mathcal{L}_\xi\bm{T}$ vanishes for all $\xi$. 

\subsubsection{The perturbation to the Riemann is a gauge independent wave}\label{sec:gwPert}

Following Isaacson \cite{isaacson:1968:gri}, consider a background spacetime with curvature radius $\mathcal{O}(L)$ a small (order $\epsilon$) amplitude gravitational wave perturbation with curvature radius $\mathcal{O}(\omega^{-1})$: i.e., in abstract index notation \cite[sec.~2.4]{wald:1984:gr}, 
\begin{subequations}\label{eq:expansion}
\begin{align}
\varphi_\epsilon^*\widetilde{R}_{abcd} &= R_{abcd} + \epsilon R^{(1)}_{abcd} + \mathcal{O}(\epsilon^2)
\end{align}
where 
\begin{align}
R_{abcd} &\sim L^{-2}\\
R^{(1)}_{abcd} &\sim \omega^2.
\end{align}
\end{subequations}
The perturbation of the metric connection $\Delta_\epsilon^*[\widetilde{\Gamma}\indices{^{a}_{be}}]$ uses its definition in terms of the metric tensor: 
\begin{subequations}
\begin{align}
\Delta_{\epsilon}^*\left[\widetilde{\Gamma}\indices{^{a}_{be}}\right] &= 
\Gamma\indices{^{a}_{be}} + 
\epsilon\Gamma\indices{^{(1)a}_{be}} + \mathcal{O}(\epsilon^2)
\end{align}
which scale with $L$ and $\omega$ as
\begin{align}
\Gamma\indices{^{a}_{be}} &\sim L^{-1},&\Gamma\indices{^{a}_{be,\delta}} &\sim L^{-2}\\
\Gamma\indices{^{(1)a}_{be}} &\sim \omega,&\Gamma\indices{^{(1)a}_{be,\delta}} &\sim \omega^{2}.
\end{align}
\end{subequations}
Corresponding to our desire to interpret the perturbation as a gravitational wave we suppose that the perturbation wavelength is very much less than the scale on which the background varies: i.e., $\delta = (\omega L)^{-1}\ll1$. The ratio of the magnitude of $\bm{R}^{(1)}$ to $\bm{R}^{(0)}$ is then $\delta^{-2}\gg1$. 

Now consider an order $\epsilon \ll 1$ coordinate gauge transformation generated by the vector field $\epsilon\xi^a$. The difference between the perturbed Riemann curvature in the old and the new gauge is given by the Lie derivative of $\bm{R}$ with respect to $\bm{\xi}$. The order $\epsilon$ gauge terms are
\begin{align}
{R}\indices{^{(1)}_{abcd}}-\bar{R}\indices{^{(1)}_{abcd}}&=
\mathcal{L}\indices{_{\bm{\xi}}}
R\indices{_{abcd}}\nonumber\\
&=
R\indices{_{abcd|f}}\xi\indices{^{f}} +    
R\indices{_{fbcd}}\xi\indices{^{f}_{|a}} + \nonumber\\
&\qquad{}
R\indices{_{afcd}}\xi\indices{^{f}_{|b}} + 
R\indices{_{abfd}}\xi\indices{^{f}_{|c}} + 
R\indices{_{abcf}}\xi\indices{^{f}_{|d}}
\end{align}
where an overbar represents a quantity in the new gauge and covariant derivatives with respect to the background metric are denoted by a vertical bar ($|$). 
The generator of the gauge transformation is independent of $\epsilon$ and $\delta$; so
\begin{subequations}
\begin{align}
\xi\indices{^a}&\sim\mathcal{O}(1)\\
\xi\indices{^a_{,b}}&\sim\mathcal{O}(1). 
\end{align}
\end{subequations}
Correspondingly, the gauge terms are $\mathcal{O}(L^{-3})$, which is  
$\mathcal{O}(L^{-1}\delta^2)$ relative to 
$R^{(1)}_{abcd}$. We conclude, with Isaacson \cite{isaacson:1968:gri}, that as long as 
$\delta<1$ the first-order perturbation to the Riemann 
$R^{(1)}_{abcd}$ (and, similarly, the Ricci 
$R^{(1)}_{ab}$ and Ricci scalar 
$R^{(1)}$) is physical: i.e., gauge independent. 

The Riemann tensor satisfies a non-linear wave equation that, in vacuum, may be written 
\begin{subequations}
\begin{align}
0 &= R\indices{_{bcde;a}^{a}} 
+S\indices{_{bcde}}\,,
\end{align}
where
\begin{align}
S\indices{_{bcde}} &= 
R\indices{^a_{cb}^f}R\indices{_{afde}} +
\left[
R\indices{^a_{cd}^f}R\indices{_{abfe}} -
R\indices{^a_{ce}^f}R\indices{_{abfd}}
\right]\nonumber\\
&\qquad{}-
R\indices{^a_{bc}^f}R\indices{_{afde}} -
\left[
R\indices{^a_{bd}^f}R\indices{_{acfe}} -
R\indices{^a_{be}^f}R\indices{_{acfd}}
\right]
\end{align}
\end{subequations}
is quadratic in the Riemann \cite[Eq.~6.9-10]{choquet-bruhat:2009:gra}. Focusing on the $\mathcal{O}(\epsilon)$ terms in this equation we have 
\begin{subequations}
\begin{align}
\nabla\indices{^a}\gv{a}R\indices*{^{(1)}_{bcde}} 
&\sim \mathcal{O}(\omega^4)\\
S\indices*{^{(1)}_{bcde}} &\sim \mathcal{O}(\omega^4\delta^2).
\end{align}
\end{subequations}
Thus, for $\delta^2\ll1$ the perturbation satisfies the vacuum wave equation on the background. 

We conclude that as long as $\delta^2\ll1$ --- i.e., the curvature radius of the perturbation is small compared to the curvature scale of the background --- the first-order perturbation to the Riemann is a gauge independent wave on the background described by the unperturbed spacetime.

\subsection{Antenna response to gravitational waves}\label{sec:gwResp}
Having established that short wave perturbations to the Riemann are gauge independent and satisfy the wave equation on the background spacetime we ask how, for short-wave perturbations, the observed clock phase is affected by the presence of gravitational waves. 

As found in section \ref{sec:diffEq} the observed clock phase is given by (see Eq.~\ref{eq:RiemannResponse})
\begin{align}\label{eq:RiemannResponseRecap}
\frac{d^2\phi}{d\tau^2} &= 
\left(\frac{d\phi/d\tau}{d\phi/dt}\right)^2
\frac{d^2\phi}{dt^2}
-
\frac{d\phi/d\tau}{\plambda\cdot\fv{U}}\left\{
\left(\frac{d\phi/d\tau}{d\phi/dt}\right)^2
\left.\left(\plambda\cdot\fv{a}_C\right)\right|_{\lambda=0}
- \left.\left(\plambda\cdot\fv{a}_R\right)\right|_{\lambda=\lambda_R}
\vphantom{
\left(\frac{d\phi/d\tau}{d\phi/dt}\right)^2
\left.\left(\plambda\cdot\fv{a}_C\right)\right|_{\lambda=0}
- \left.\left(\plambda\cdot\fv{a}_R\right)\right|_{\lambda=\lambda_R}
-\left(\frac{d\phi}{/d\tau}\right)^2
\int_0^{\lambda_{R}}
\left[\fv{K}^2-\fv{R}(\pphi,\plambda,\pphi,\plambda)\right]d\lambda
}
\right.
\nonumber\\
&\qquad\qquad
\left.
\vphantom{
\left(\frac{d\phi/d\tau}{d\phi/dt}\right)^2
\left.\left(\plambda\cdot\fv{a}_C\right)\right|_{\lambda=0}
- \left.\left(\plambda\cdot\fv{a}_R\right)\right|_{\lambda=\lambda_R}
-\left(\frac{d\phi}{/d\tau}\right)^2
\int_0^{\lambda_{R}}
\left[\fv{K}^2-\fv{R}(\pphi,\plambda,\pphi,\plambda)\right]d\lambda
}
-\left(\frac{d\phi}{d\tau}\right)^2
\int_0^{\lambda_{R}}
\left[\fv{K}^2-\fv{R}(\pphi,\plambda,\pphi,\plambda)\right]d\lambda
\right\}\,
\end{align}
where $\bm{a}_R$ and $\bm{a}_C$ are the 4-accelerations of the receiver/observer and the clock, $\plambda$ is the null-geodesic field connecting the clock and receiver/observer, and $t$ and $\tau$ are the clock and receiver/observer proper times as measured at either end of the connecting null geodesics. Of the terms on the right-hand side of Equation~(\ref{eq:RiemannResponseRecap}) the gravitational wave perturbation does not affect $d\phi/dt$, $d^2\phi/dt^2$, which are intrinsic properties of the clock measured in the clock's proper reference frame. The remaining terms --- $d\phi/d\tau$, $\bm{a}_R$, $\bm{a}_C$, $\plambda$, $\bm{K}$, $\lambda_R$, and $\bm{R}$ --- are all perturbed and contribute to the perturbation of $d^2\phi/d\tau^2$.  In Appendix \ref{sec:scaling} we evaluate the scaling of all the contributions to the perturbation of Equation \ref{eq:RiemannResponseRecap} with respect to the gravitational wave's amplitude (described by $\epsilon \ll 1$), its angular frequency ($\omega$) and the ratio of the radiation wavelength to the background curvature scale ($\delta = (\omega L)^{-1}$).  At first order in $\epsilon$ the terms in Equation \ref{eq:RiemannResponseRecap} scale as
\begin{subequations}\label{eq:scalings}
\begin{align}
\pert{\left(\frac{d\phi/d\tau}{d\phi/dt}\right)^2
\frac{d^2\phi}{dt^2}} &\sim \mathcal{O}(1)\\
\pert{\frac{d\phi/dt}{\plambda\cdot\fv{U}}
\left(\frac{d\phi/d\tau}{d\phi/dt}\right)^2
\left.\left(\plambda\cdot\fv{a}_C\right)\right|_{\lambda=0}} &\sim\mathcal{O}(1)\\
\pert{\frac{d\phi/dt}{\plambda\cdot\fv{U}}\left.\left(\plambda\cdot\fv{a}_R\right)\right|_{\lambda=\lambda_R}} &\sim \mathcal{O}(1)\\
\pert{\frac{d\phi/dt}{\plambda\cdot\fv{U}}\left(\frac{d\phi}{d\tau}\right)^2
\int_0^{\lambda_{R}}
\fv{K}^2 d\lambda} &\sim \mathcal{O}(\omega^2\delta)\\
\pert{\frac{d\phi/dt}{\plambda\cdot\fv{U}}\left(\frac{d\phi}{d\tau}\right)^2
\int_0^{\lambda_{R}}
\fv{R}(\pphi,\plambda,\pphi,\plambda)d\lambda} &\sim\mathcal{O}(\omega^2). 
\end{align}
\end{subequations}

Thus we find our principal result: in the limit of high frequency gravitational waves (i.e., $\delta\ll1$: the radiation wavelength is much smaller than the background curvature scale) the contribution to the observed evolution of the clock phase $d^2\phi/d\tau^2$ owing to the gravitational waves is given by the gravitational wave contribution to the Riemann curvature integrated along the unperturbed null geodesic trajectory linking the clock and receiver/observer, or 
\begin{equation}\label{eq:ResponsePerturbation}
\left(\frac{d^2\phi}{d\tau^2}\right)^{(1)} = \frac{d\phi/dt}{\plambda\cdot\fv{U}}\left(\frac{d\phi}{d\tau}\right)^2
\int_0^{\lambda_{R}}
\fv{R}^{(1)}(\pphi,\plambda,\pphi,\plambda)d\lambda + \mathcal{O}(\delta)
\end{equation}
where $R^{(1)}$ is the gravitational wave contribution to the Riemann curvature and all other terms correspond to their unperturbed quantities.

%\input{discussion.tex}
%!TEX root = ./ms.tex
% $Id: discussion.tex 6959 2013-10-10 02:37:44Z mkoop $
\section{Discussion}\label{sec:discussion}
\subsection{Gauge invariant response}
Our expression for the observed Doppler,
\begin{align}\label{eq:dopplerDisc}
\frac{d^2\phi}{d\tau^2} &= 
\left(\frac{d\phi/d\tau}{d\phi/dt}\right)^2
\frac{d^2\phi}{dt^2}
-
\frac{d\phi/d\tau}{\plambda\cdot\fv{U}}\left\{
\left(\frac{d\phi/d\tau}{d\phi/dt}\right)^2
\left.\left(\plambda\cdot\fv{a}_C\right)\right|_{\lambda=0}
- \left.\left(\plambda\cdot\fv{a}_R\right)\right|_{\lambda=\lambda_R}
\vphantom{
\left(\frac{d\phi/d\tau}{d\phi/dt}\right)^2
\left.\left(\plambda\cdot\fv{a}_C\right)\right|_{\lambda=0}
- \left.\left(\plambda\cdot\fv{a}_R\right)\right|_{\lambda=\lambda_R}
-\left(\frac{d\phi}{/d\tau}\right)^2
\int_0^{\lambda_{R}}
\left[\fv{K}^2-\fv{R}(\pphi,\plambda,\pphi,\plambda)\right]d\lambda
}
\right.
\nonumber\\
&\qquad\qquad
\left.
\vphantom{
\left(\frac{d\phi/d\tau}{d\phi/dt}\right)^2
\left.\left(\plambda\cdot\fv{a}_C\right)\right|_{\lambda=0}
- \left.\left(\plambda\cdot\fv{a}_R\right)\right|_{\lambda=\lambda_R}
-\left(\frac{d\phi}{/d\tau}\right)^2
\int_0^{\lambda_{R}}
\left[\fv{K}^2-\fv{R}(\pphi,\plambda,\pphi,\plambda)\right]d\lambda
}
-\left(\frac{d\phi}{d\tau}\right)^2
\int_0^{\lambda_{R}}
\left[\fv{K}^2-\fv{R}(\pphi,\plambda,\pphi,\plambda)\right]d\lambda
\right\}\,,
\end{align}
(see Eq.~\ref{eq:RiemannResponse})
is manifestly gauge-invariant. Each term on the right hand side of Equation \ref{eq:dopplerDisc} is either an intrinsic property of the clock ($d\phi/dt$, $d^2\phi/dt^2$) or a spacetime scalar. Equation \ref{eq:dopplerDisc} was derived entirely in terms of the physical measurements made at the receiver/observer or at the clock. At no point was it necessary to identify special coordinates, fix a gauge, or introduce privileged observers. Furthermore, each term is composed from quantities that are physical observables or, equivalently, well-defined spacetime properties:
\begin{itemize}
\item $\fv{a}_C$ and $\fv{a}_R$ are the clock and receiver/observer 4-acceleration, just as would be measured by an inertial accelerometer co-located with the clock or the receiver;
\item $\plambda$ is the tangent to the null geodesic that connects clock and receiver; 
\item $\pphi$ is defined everywhere on the clock world line in terms of the clock 4-velocity and intrinsic frequency (see Eq.~\ref{eq:bcPhi}), and along each of the null geodesic connecting the clock and the receiver/observer via the equation of geodesic deviation;
\item $\fv{K}$ is defined everywhere along the receiver/observer worldline in terms of the receiver's 4-acceleration and the clock's observed proper motion (see Eq.~\ref{eq:bcLambda}) and then everywhere along the null geodesics via the equation of geodesic deviation; and
\item $\fv{R}$ is the spacetime Riemann curvature. 
\end{itemize}

Finally, when the gravitational radiation wavelength is short compared to the background curvature lengthscale, the gravitational wave contribution to the Riemann curvature is a physical, gauge invariant property of spacetime. Under these conditions Equation \ref{eq:ResponsePerturbation} gives the leading contribution to the observed Doppler,
\begin{equation} \label{eq:dopplerPertDisc}
\left(\frac{d^2\phi}{d\tau^2}\right)^{(1)} = \frac{d\phi/dt}{\plambda\cdot\fv{U}}\left(\frac{d\phi}{d\tau}\right)^2
\int_0^{\lambda_{R}}
\fv{R}^{(1)}(\pphi,\plambda,\pphi,\plambda)d\lambda + \mathcal{O}(\delta),
\end{equation}
where $\fv{R}^{(1)}$ is the gravitational wave contribution to the Riemann curvature and the integration is along the unperturbed null geodesic trajectories of the background spacetime.

\subsection{Physical interpretation}
Each of the terms on the right hand side of Equation \ref{eq:dopplerDisc} also has a clean, unambiguous physical interpretation. The first term, involving $d^2\phi/dt^2$, accounts for the intrinsic evolution of the clock's frequency. The second and third terms are the contributions owing to the relativistic generalization of the classical Doppler effect associated with the receiver/observer acceleration along the path from clock to receiver/observer. The remaining two terms, involving integrals along the null geodesic from the clock to the receiver, describe the light-time contributions to the Doppler. Recalling (Eq.~\ref{eq:K})
\begin{align}\label{eq:Kdisc}
\fv{K} &= \gv{\pphi}\plambda,
\end{align} 
the term in $\fv{K}^2$ accounts for what we term \emph{kinematic} light-time contributions: kinematic because they are associated with the spacetime null path evolution without regard to the cause behind the path evolution. Finally, the term involving the integral of $\fv{R}(\pphi,\plambda,\pphi,\plambda)$ is the light-time contribution owing to the effects of general relativistic gravity along the path. 

Since the contribution from the integral of $\fv{R}(\pphi,\plambda,\pphi,\plambda)$ along the null geodesics is the only contribution that vanishes when ``gravity'' vanishes (i.e., spacetime is flat) and, as was shown in Section \ref{sec:scaling}, dominates the gravitational wave contribution when waves are present, it may be fairly said to encapsulate the unique contribution of relativistic gravity to the observed Doppler. 

\subsection{Application to pulsar timing}
It is instructive to compare our expression for the observed clock Doppler to the timing model used today for pulsar timing. In that model \cite{edwards:2006:tnp} the time-varying contributions to the expected pulse arrival times are identified individually and, for relativistic contributions, computed perturbatively in a combination of specific coordinate systems that are related to each other through appropriate coordinate transformations \cite{damour:1986:grc,Stairs:2003gc,2006MNRAS.369..655H,Edwards:2006in}. All effects --- relativistic and non-relativisitc --- are grouped together in a small set of high-level ``delays'' (sometimes ``corrections'') of common conceptual or perturbative origin. The delays associated with the vacuum propagation of the electromagnetic pulse are referred to as the R{\o}mer, Einstein and Shapiro delays \cite{1986AIHS...44..263D,2006MNRAS.369..655H}. The R{\o}mer delay terms are strictly ``geometric'' --- e.g., the light travel time across the solar system (more precisely, between the pulse arrival at the solar system barycentre and the observatory) --- without taking into account any relativistic effects. The R{\o}mer Delay terms also includes light-time corrections owing to pulsar proper motion (i.e., the Shklovskii effect \cite{shklovskii:1970:pco}), corrections owing to parallax, and corrections owing to aberration (associated with Earth's transverse velocity). The Einstein delay accounts for the special-relativistic time-dilation and gravitational redshift effects as Earth moves in its orbit and through the gravitational potential of Sol and Luna. Finally, the Shapiro delay describes additional delays owing to the Shapiro effect \cite{1964PhRvL..13..789S} associated with Sol and other solar system bodies. (If the pulsar is part of a binary system there are an additional three terms corresponding to these same effects in the pulsar's binary system.) 

When Equation \ref{eq:dopplerDisc} is applied to pulsar timing the world lines describing the pulsar and Earth's spacetime trajectories are geodesic; so, ignoring the displacement of the observatory from Earth's worldline, the acceleration terms in Equation~\ref{eq:dopplerDisc} vanish. Of the three remaining terms, one is due to the pulsar's period evolution evaluated at the retarded time and adjusted by $(dt/d\tau)^2$, which includes some of what are presently computed as the Einstein delay corrections. Recalling Equation \ref{eq:Kdisc} for $\fv{K}$, the integral of $\fv{K}^2$ along the pulsar-Earth null geodesic accounts for contributions owing to the change in the null geodesic path as a function of receiver/observer time: e.g., the R{\o}mer delay and the Shklovskii effect \cite{shklovskii:1970:pco}. Writing $d\phi/d\tau$ as $(d\phi/dt)(dt/d\tau)$ we see that the Einstein delay corrections to these terms are already included. Finally, the Shapiro delay (and, of course, gravitational wave) corrections are part of the integral of $\fv{R}(\pphi,\plambda,\pphi,\plambda)$ over the null geodesic. A fuller discussion of how Equation \ref{eq:dopplerDisc} leads to a pulsar timing formula is part of a work in progress. 

\subsection{Gravitational wave detection when $\delta \gtrsim1$}
Pulsar timing for gravitational wave detection is generally aimed at detecting gravitational waves with periods on order years \cite{jenet:2009:nan}. In Earth's neighborhood the background curvature scale owing to Sol is on order two light-months; in the neighborhood of the pulsar the curvature scale is much smaller. For pulsar timing, then, the condition $(\omega L)^{-1}=\delta\ll1$, which is required to distinguish gravitational waves and background, does not hold near Earth or near the pulsar. 

Nevertheless, in the case of pulsar timing the failure of the condition $\delta\ll1$ near Earth (or the pulsar) is not significant. The key to understanding how this is so is to recognize that the gravitational wave response is contributed by the \emph{integral of} $\fv{R}(\pphi,\plambda,\pphi,\plambda)$ \emph{along the null geodesic from the pulsar to the Earth-bound observatory.} Over most of that path $\delta\ll1$: i.e., it is over only a very small fraction of that path near, near its endpoints, that there is any ambiguity over what is wave and what is background. 

Since the mass of Sol and the mass of the pulsar are of the same order the spacetime curvature at the same distance from each is also of the same order. Without loss of generality, focus attention on the importance of the spacetime curvature owing to Sol in Earth's neighborhood. The background curvature scale a distance $r$ from Sol is $\sim\text{M}_\odot/r^3$; so, 
\begin{align}
\delta^2 &\simeq \frac{\text{M}_\odot}{\omega^2 r^3}\,.
\end{align}
If we require $\delta^2<10^{-2}$ then it is only at distances
\begin{align}
r&\lesssim 
5\,\text{au}\left({1+\cos\iota}\right)^{-1/3}
\left(\frac{f}{\text{yr}^{-1}}\right)^{-2/3},
\end{align}
where $\iota$ is the angle between the wave propagation direction and the Earth-pulsar line-of-sight, that we need be concerned. (For waves propagating with $\iota\simeq\pi$ this is large; however, $\fv{R}(\pphi,\plambda,\pphi,\plambda)\sim\sin^2\iota$; so, when  $r$ might be large the gravitational wave contribution vanishes.) This corresponds to a fraction 
\begin{align}
\frac{\omega r}{2\pi}&\simeq
1.6\times10^{-5}
\left({1+\cos\iota}\right)^{-1/3}
\left(
\frac{f}{\text{yr}^{-1}}
\right)^{2/3}
\end{align}
of a radiation wavelength and a similarly small fraction of the contribution to the integral of $\fv{R}(\pphi,\plambda,\pphi,\plambda)$. We conclude that the failure to satisfy $\delta\ll1$ near either Earth or the pulsar does not significantly affect the use of pulsar timing for gravitational wave detection with wavelengths greater than the spacetime curvature in the solar system.

Do similar problems arise when we treat gravitational waves as (TT gauge) metric perturbation? They do, but in those treatments the problems have been overlooked. In conventional treatments the gravitational wave contribution to pulsar arrival times is proportional to the difference of the anti-derivative of what is supposed to be the TT gauge metric perturbation at Earth and the retarded-time anti-derivative at the pulsar. We say ``supposed'' because the metric perturbations at Earth and at the pulsar must be computed as if spacetime at Earth  the pulsar is flat: i.e., that the pulsar, Earth, Sol and the solar system have no effect whatsoever on the metric perturbation representation of the gravitational wave perturbation. This is far from the case, especially near the pulsar. Reviewing Equation \ref{eq:dopplerDisc}, however, we see that the conventional result comes about when $\delta\ll1$ over the null geodesic connecting the pulsar and Earth except for a small (relative to the gravitational radiation wavelength) portion of the path near its endpoints. 

\subsection{Application to interferometric detectors}
\begin{figure}
\includegraphics[width=0.5\textwidth]{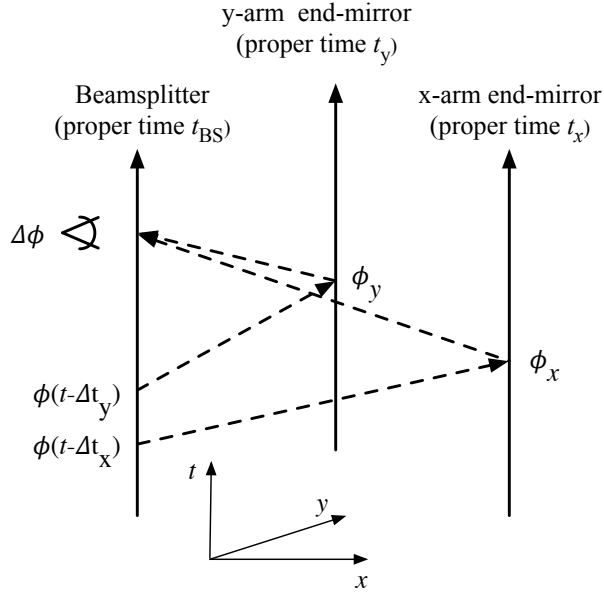}
\caption{A schematic spacetime diagram describing a single-pass delay-line interferometer used to detect gravitational waves. The worldlines of the beamsplitter, $x$-arm and $y$-arms end-mirrors are shown as vertical, solid lines. The receiver/observer at the interferometer output port records the phase difference in the light arriving along the null paths, represented by the dotted lines. Each of the end-mirrors may be considered its own clock, slaved to the laser illuminating the input port of the interferometer. To describe this interferometric detector use Equation \ref{eq:dopplerDisc} to determine $\phi_x(t_{x})$ and $\phi_y(t_y)$ on the respective end-mirror worldlines, and then use Equation \ref{eq:dopplerDisc} again to determine $\Delta\phi(t_{\text{BS}})$ how these are observed back at the beamsplitter.}\label{fig:ifo}
\end{figure}

Equation \ref{eq:dopplerDisc} may also be applied to find a gauge invariant description of the response of laser interferometric gravitational wave detector. Referring to Figure \ref{fig:ifo}, which shows in schematic a single-pass delay line interferometer, the interferometer's response is the phase difference $\Delta\phi$ of the clock's image reflected from the $x$-arm beamsplitter ($\phi_x$) and the $y$-arm beamsplitter ($\phi_y$). Equation \ref{eq:dopplerDisc} describes how the phases $\phi_x$ and $\phi_y$ are observed at the output port. It also describes how $\phi_x$ and $\phi_y$ are determined from the light phase at the interferometer input port. The response of an interferometric detector thus follows from the iterative use of Equation \ref{eq:dopplerDisc}. A detailed discusson of interferometric detector response is part of work in progress. 

\subsection{Relationship to metric perturbation treatment of gravitational waves}
In an arbitrary gauge the $\mathcal{O}(\epsilon)$ perturbation to the Riemann curvature can be expressed in terms of the metric perturbation $h\indices{_{ab}}$ as
\begin{align}
R\indices*{^{(1)}_{abcd}} &= 
h\indices{_{ed}}R\indices*{^{(0)}_{abc}^e} -
\frac{1}{2}
\left(
h\indices{_{cd|ba}} +
h\indices{_{bd|ca}} -
h\indices{_{bc|da}} -
h\indices{_{cd|ab}} -
h\indices{_{ad|cb}} +
h\indices{_{ac|db}}
\right)
\end{align}
where, as before, latin indices $a$, $b$, $c$, $d$ and $e$ denote abstract indices. 

In Minkowski space the background curvature $R\indices*{^{(0)}_{abcd}}$ vanishes. If, as is customary, we assume that in the Minkowski background global Lorentz frames the receiver/observer is undergoing force-free motion, then we may introduce TT coordinates in the Lorentz Frame in which the receiver/observer is at rest. In those coordinates $\pphi\propto\partial_t$ and the coordinate components of the Riemann relevant for determining the gravitational wave response are 
\begin{align}\label{eq:TTGauge}
R\indices{_{t\mu t\nu}} &= -\frac{1}{2}\frac{\partial^2h\indices{_{\mu\nu}}}{\partial t^2},
\end{align}
where the indices in Equation \ref{eq:TTGauge} denote coordinate components. Thus, when TT gauge exists we recover the usual representation of gravitational waves in terms of the TT gauge metric perturbation. In the view taken here, however, this association validates the TT-gauge description of gravitational waves and not the other way around. 

An advantage of our gauge-independent treatment is that it applies even in environments where special gauges --- e.g., TT gauge --- do not exist.

\subsection{Numerical relativity}
Numerical relativity exploits the coordinate gauge freedom to place the Einstein equations in forms that allow for stable evolution and the use of standard solution techniques (e.g., ``black-box'' elliptic, parabolic of first-order hyperbolic system solvers) \cite{lehner:2001:nra}. To convert the results of these simulations to a TT-gauge metric perturbation is a non-trivial excercise that introduces additional, non-trivial numerical errors; however, a direct product of numerical relativity simulations is the connection and the Riemann tensor, from which the response of laser interferometric, spacecraft Doppler tracking or pulsar timing gravitational wave detector response can be determined without the need for solving the gauge equations necessary to transform the simulation results into TT gauge. Expressing the detector response in terms of the gauge independent Riemann, as opposed to the metric perturbation in a specific gauge, eliminates unecessary steps and associated numerical errors when using numerical relativity simulations to aid in the analysis or interpretation of gravitational wave observations.

\subsection{Misconceptions regarding the operation of gravitational wave detectors}\label{sec:misconceptions}
With Equations \ref{eq:dopplerDisc}, \ref{eq:dopplerPertDisc} and the intermediate results summarized in Appendix \ref{sec:scaling} we are in a position to address the mistatements and misconceptions, highlighed in Section \ref{sec:intro}, that have appeared in the pedagogical and research literature regarding how gravitational wave detectors function and what they respond to. As is common in these discussions we specialize to a Minkowski background with the clock and receiver/observer at relative rest in the background. In this case Equation \ref{eq:dopplerDisc} and \ref{eq:dopplerPertDisc} together reduce to 
\begin{align}\label{eq:dopplerMink}
\frac{d^2\phi}{d\tau^2} &= 
-\frac{\left(d\phi/d\tau\right)^3}{\plambda\cdot\fv{U}}
\int_0^{\lambda_{R}}
\fv{R}^{(1)}(\pphi,\plambda,\pphi,\plambda)d\lambda,
\end{align}
where the integration path is along the appropriate (piecewise) null geodesics of the background spacetime. 

An important property of this result is its agreement with the Einstein Equivalence Principle \cite{will:1992:cbg}, which asserts that true gravitational effects are discernible only in experiments of sufficiently large scale. This should be no less the case for gravitational waves than for all other gravitational phenomena. In Equation \ref{eq:dopplerMink} we see that the magnitude of the gravitational wave effect depends directly on the length of the path traversed by the electromagnetic wave. Viewed from the perspective of gravitational wave detection via pulsar timing (or spacecraft Doppler tracking), the gravitational wave contribution to the observed pulse arrival time at Earth depends on the Riemann curvature along the path traversed by the electromagnetic signal from the distant pulsar: i.e., contrary to the shibboleth \cite[e.g.,]{Cordes:2012cq} that pulsar timing measures the effect of gravitational waves near Earth and the pulsar, \emph{the spacetime properties in the immediate neighborhood of the Earth and the pulsar are irrelevant for the detection of gravitational waves via pulsar timing.} 

\begin{figure}
\includegraphics[width=0.5\textwidth]{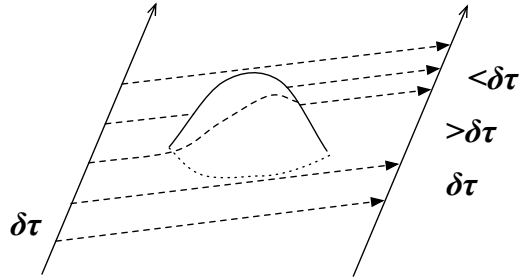}
\caption{Gravitational wave detectors measure a projection of the spacetime curvature along paths that link the detector's components. See Section \ref{sec:misconceptions} for discussion.}\label{fig:curvature}
\end{figure}

For a graphical illustration of this latter point, refer to Figure \ref{fig:curvature}, which shows the worldlines of a clock and a receiver, propagating freely in a spacetime that is flat in their respective neighborhoods. A ``lump'' of curvature arises between them, persists for some time, and then evaporates. As shown, the curvature leads to variations in the pulse arrival times at the receiver even though spacetime in the neighborhood of the clock and receiver is always flat and described by the Minkowskii metric: i.e., there is no ``metric perturbation'' on either worldline. 

We conclude that it is gravitational wave induced difference in light travel time as it propagates along the unperturbed null trajectories connecting the clock and observer that leads to the detector response. There is no meaningful or observable sense in which the gravitational waves accelerate or move the detector components (cf.~\cite{Saulson:2006wg,thorne:2002:gwa}); no significant contribution to the resonse that arises from the deflection of the electromagnetic waves by the gravitational waves (cf.~\cite{anholm:2009:osf}); the Riemann curvature tensor does not influence or affect light in any special way (cf.~\cite{thorne:2002:gwa}). Finally, red/blue-shifting of the light in a detector depends only on the choice of observer and can, by appropriate choice of observer, be made to vanish (cf.~\cite{1997AmJPh..65..501S}). 

\subsection{Use of geodesic deviation}\label{sec:geodev}
The equation of geodesic deviation arises in our analysis because the vector $\pphi$ is the deviation vector associated with the null geodesic congruence $\plambda$ along which the phase record of the clock is propagated to the observer (see sec.~\ref{sec:doppler}). Other discussions of detector response also make use of the geodesic deviation equation \cite[e.g.,]{schutz:1990:fci,faraoni:2007:cma}. What makes the discussion and result here unique? The starting point for the discussion here is the physical measurement made at the receiver observer. The analysis that follows is fully relativistic, invokes no special coordinate systems, gauges, or privileged observers, and ends with a fully general result that involves only the intrinsic properties of the clock, the receiver/observer, their respective worldlines, and the spacetime geometry. The result is valid independent of the detector size relative to the radiation wavelength and even when a distinction between the gravitational radiation and the background curvature is ambiguous. In the analysis presented here the equation of geodesic deviation arises as a matter of course and not as a fundamental precept. 

In contrast, \citet{schutz:1990:fci,faraoni:2007:cma} begin by asserting that the geodesic deviation equation describes the instantaneous spatial separation (in some ill-defined detector ``rest-frame'') between the coordinate stationary world lines of the clock (laser/beamsplitter) and the receiver/observer (end mirror), which may not be geodesics. The spatial separation divided by the light-speed $c$, is then supposed to give the light travel time from clock to observer. Finally, the observed phase is calculated under the assumption that the gravitational radiation wavelength is much longer than the detector size or light-storage time. In these descriptions geodesic deviation is applied heuristically to characterize a coordinate dependent non-deviation vector between two non-geodesics. 

Previous analyses that depend upon or involve in some way the geodesic deviation equation, while doubtless having some heuristic value, offer, at best, results valid only in very special circumstances: a Minkowski background, detectors small compared to a radiation wavelength and with light storage times short compared to the radiation period. They are not generalizable to backgrounds that are not Minkowski, detectors that approach or exceed a radiation wavelength (e.g., LISA, spacecraft Doppler tracking or pulsar timing), or detectors whose light storage time approaches or exceeds a radiation period (e.g., LIGO and kHz frequency gravitational waves associated with neutron star binary coalescence). Finally, these derivations, through their focus on special coordinate systems where clock and receiver position and spatial separation can be said to play an important role, suggest that, in some way, gravitational radiation violates the equivalence principle and can be measured at a point. In this way the heuristic picture they offer is not that different than one focused upon metric perturbations. 

%\input{conclusion.tex}
%!TEX root = ./ms.tex
% $Id: conclusion.tex 6960 2013-10-10 14:32:27Z lsfinn $
\section{Conclusion}\label{sec:conclusions}
Most modern gravitational wave detectors --- including laser interferometers, pulsar timing arrays, spacecraft doppler tracking --- rely upon the waves effect on light travel time to detect the wave's presence. The response of these detectors is conventionally expressed in terms of a metric perturbation description of the gravitational waves. Metric perturbations, however, are coordinate gauge dependent quantities. Their use obscures the physical nature of the gravitational wave phenomena and the principles that governs the operation of these detectors \cite{faraoni:1992gu,saulson:1997:ilw,garfinkle:2006:gia,faraoni:2007:cma,finn:2009:roi,kennefick:2007:tas} and lead to ``explanations'' of detector function that are at odds with the equivalence principle, which is the most basic assertion of general relativity. 

The response of modern gravitational wave detectors may also be expressed in terms of their interaction with spacetime curvature, which is the physically unambiguous property of spacetime that, in general relativity, underlies all gravitational phenomena. 
Such an expression of the detector response, and each term that comprises it, is separately gauge invariant and has a clear physical interpretation. The expression for the response is manifestly consistent with the equivalence principle: i.e., the role of all gravitational phenonmena in the response --- wave or otherwise --- clearly involves measurements made over finite spacetime intervals. The gravitational wave contribution to the response is separately identifiable and the way in which it leads to light-time fluctuations in the detector avoids all the suspect analogies (moving mirrors, redshifting and blueshifting of light, etc.) that plague descriptions of detector functioning based upon metric perturbations. Contributions from other physical phenomena also appear naturally: e.g., in the case of pulsar timing, all geometrical and kinematical effects (R{\o}mer delay, parallax, aberration, Shklovskii effect \cite{shklovskii:1970:pco}), special relativistic and general relativistic effects (special and general relativistic Einstein delay, Shapiro time delay and gravitational wave contributions) all appear naturally. 

The ``curvature-based'' response function describes all light-time based detectors in all regimes: i.e., terrestrial laser interferometric detectors, whose physical dimensions and light storage times are shorter than the typical radiation wavelength \cite{harry:2010:aln,accadia:2010:sap,rakhmanov:2008:hct}; proposed space-based detectors based upon the LISA heritage, where the detector size and light storage time is on-order the radiation wavelength \cite{stebbins:2009:rl,Danzmann:2011tn}; and spacecraft doppler and pulsar timing, where the radiation wavelength is very much shorter than the detector size or light storage time \cite{hellings:1981:sgd,armstrong:2006:lgw,foster:1990:cpt,hobbs:2010:ipt}. It meshes well with the use of numerical relativity calculation to determine detector response: i.e., it does not require extracting from these calculations gauge dependent quantities that are not directly calculated in order to compute the director response. 

In companions papers we shall explore in greater depth how expressing the gravitational wave response in terms of the Riemann curvature can be applied to delay-line and Fabrey-Perot interferometric detectors; the role of non-uniform motion (e.g., rotation) in LISA-heritage detectors; the effects of micro-lensing in pulsar timing gravitational wave detection \cite{Pshirkov:2008de,Dai:2010iq}; and the use of the Riemann curvature response function to provide an alternative derivation of a general-relativistic pulsar timing formula. 

Sometime in the next decade gravitational wave observations will begin to constrain our understanding of astronomical phenomena: when that moment arrives, gravitational wave astronomy will be born. The acceptance of gravitational wave observations as a tool of astronomical discovery can only be hastened by an intuitive, simple and physically correct understanding of gravitational waves and their detection: what they are, how they are generated, how they propagate, and how they interact with a detector. Expressing the gravitational wave detector response in terms of the Riemann curvature is but one step in that direction. 

\begin{acknowledgments}
LSF thanks the Aspen Center for Physics for their hospitality and thanks David Garfinkle and John Baker for discussions. This work was supported by National Science Foundation Grant Numbers 09-40924 and 09-69857 awarded to The Pennsylvania State University and supported by the Natural Science and Research Council of Canada. 
\end{acknowledgments}

%\input{appendix}
%!TEX root = ./ms.tex
% $Id: appendix.tex 6938 2013-09-17 18:27:38Z lsfinn $

\appendix
\section{Scaling of contributions to the perturbation of Equation \ref{eq:RiemannResponse}}\label{sec:scaling}
In Section \ref{sec:gwResp} we concluded that the gravitational wave contribution to $d^2\phi/d\tau^2$ depends only on the integral of the gauge independent perturbation to the Riemann curvature along the unperturbed null geodesic connecting clock and receiver/observer. Our conclusion rests upon the scaling of the different contributions to the righthand side of Equation \ref{eq:RiemannResponseRecap} with respect to the radiation's amplitude (described by the small parameter $\epsilon$), it's angular frequency ($\omega$), and the ratio of the background curvature scale ($L$) to the radiation wavelength ($\delta=(\omega L)^{-1}$). In this appendix we establish those scalings, which we summarized in Equations \ref{eq:scalings}. 

\subsection{Perturbations of $\plambda$, $\bm{U}$ and $\bm{V}$}
Each of $\plambda$, $\bm{U}$ and $\bm{V}$ is constrained by definition: in particular, $\plambda$ is null and $\fv{U}^2=\fv{V}^2=-1$. From these constraints we immediately conclude that 
\begin{subequations}\label{eq:orderPUV}
\begin{align}
\plambda\indices{^{(0)}},\, \bm{U}\indices{^{(0)}},\, \bm{V}\indices{^{(0)}}&\sim\mathcal{O}(1)\\
\plambda\indices{^{(1)}},\, \bm{U}\indices{^{(1)}},\, \bm{V}\indices{^{(1)}}&\sim\mathcal{O}(1).
\end{align}
\end{subequations}

\subsection{Perturbation of $d\phi/d\tau$}
Recall 
\begin{align}
\frac{d\phi}{d\tau} &= \frac{\plambda\cdot\fv{U}}{\plambda\cdot\fv{V}}\frac{d\phi}{dt}. 
\end{align}
The clock is presumed to be much smaller than a radiation wavelength. Since $d\phi/dt$ is an intrinsic property of the clock it is unaffected by the perturbation. Taking advantage of Equations \ref{eq:orderPUV} we conclude
\begin{align}
\Delta^1_{\varphi}\left[\frac{d\phi}{d\tau}\right] &= \mathcal{O}(1).
\end{align}

\subsection{Perturbation of $\bm{a}\cdot\plambda$}
The accelerations $\bm{a}_R$ and $\fv{a}_C$ are the result of applied forces: i.e., they are unaffected by the gravitational wave perturbation. Taking advantage of Equations \ref{eq:orderPUV}, 
\begin{align}
\Delta^{1}_{\varphi}[\plambda\cdot\bm{a}] &\sim \mathcal{O}(1). 
\end{align}

\subsection{Perturbation of $\fv{K}^2$}
Recalling
\begin{align}
\bm{K} &= \gv{\pphi}{\plambda}
\end{align}
we conclude
\begin{subequations}
\begin{align}
\fv{K}^{(0)}&\sim\mathcal{O}(L^{-1})\\
\fv{K}^{(1)}&\sim\mathcal{O}(\omega)
\end{align}
\end{subequations}

\subsection{Perturbation of $\int_0^{\lambda_R}\left[\cdots\right]d\lambda$}

The perturbation of an integral of the form 
\begin{align}
\int_0^{\lambda_R} f(x^\mu(\lambda))\, d\lambda 
\end{align}
involves the perturbation of the integration $0$ and $\lambda_R$, the integration path $x^\mu(\lambda)$, and the scalar function $f$. In our case the integration endpoints are defined to be 
\begin{subequations}
\begin{align}
x^\mu(0) &= \mathcal{P}\\
x^\mu(\lambda_R) &= \mathcal{Q}
\end{align}
\end{subequations}
where $\mathcal{Q}$ is the point on the receiver/observer worldline along the null geodesic path $x^\mu(\lambda)$ leaving the clock worldline at $\mathcal{P}$. By definition $\lambda=0$ at $\mathcal{P}$. Denoting the perturbed path and the perturbed integration endpoint $\lambda_R$
\begin{subequations}
\begin{align}
x^\mu(\lambda) &= x^{(0)\mu}(\lambda) + \epsilon \delta x^{(1)\mu}(\lambda) + \mathcal{O}(\epsilon^2)\\
\lambda_R &= \lambda^{(0)}_{R} + \epsilon\delta\lambda^{(1)}_{R} + \mathcal{O}(\epsilon^2),  
\end{align}
\end{subequations}
write the perturbed integral as 
\begin{align}\label{eq:pertInt}
\int_0^{\lambda_R} f(x^\mu(\lambda))\, d\lambda 
&= \int_0^{\lambda^{(0)}_R} f^{(0)}(x^{(0)\mu}(\lambda))\, d\lambda \nonumber\\
&\qquad{}
+ \epsilon\int_0^{\lambda^{(0)}_R}\left[
    f^{(1)}(x^{(0)\mu}(\lambda))
  + \delta x^{(1)\mu}\frac{\partial f^{(0)}}{\partial x^\mu}(x^{(0)\mu}(\lambda))
\right]\, d\lambda\nonumber\\
&\qquad{}+\int_0^{\epsilon\delta\lambda_R^{(1)}} f^{(0)}(x^{(0)}(\lambda_R^{(0)}+\lambda))d\lambda. 
\end{align}
The last term on the right hand side of Equation \ref{eq:pertInt} may be further expressed
\begin{align}
\int_0^{\epsilon\delta\lambda_R^{(1)}} f^{(0)}(x^{(0)}(\lambda_R^{(0)}+\lambda))d\lambda &=
\epsilon\int_0^{\delta\lambda_R^{(1)}} f^{(0)}(x^{(0)}(\lambda_R^{(0)}+\epsilon\lambda))d\lambda \nonumber\\
&= \epsilon\delta\lambda_R^{(1)} 
f^{(0)}(\lambda_R^{(0)})
\end{align}
Since the perturbation $\plambda\indices{^{(1)}}$ is $\mathcal{O}(1)$ the path perturbation $\delta x^{(1)\mu}$ and the integration endpoint perturbation $\delta\lambda^{(1)}_R$ are also $\mathcal{O}(1)$. Correspondingly,  
\begin{align}
\Delta^{1}_{\varphi}\left[\int_0^{\lambda_R} f(x^\mu(\lambda))\, d\lambda\right] &=
\int_0^{\lambda^{(0)}_R}\left[
    f^{(1)}(x^{(0)\mu}(\lambda))
  + \delta x^{(1)\mu}\frac{\partial f^{(0)}}{\partial x^\mu}(x^{(0)\mu}(\lambda))
\right]\, d\lambda
+\delta\lambda_R^{(1)}f^{(0)}(\lambda_R^{(0)})
\end{align}

We have two terms to consider. For the first,
\begin{subequations}
\begin{align}
f &= \fv{K}^2\\
f^{(0)} &\sim \mathcal{O}(L^{-2})\\
\frac{\partial f^{(0)}}{\partial x^\mu}&\sim\mathcal{O}(L^{-3})\\
f^{(1)} &\sim \mathcal{O}(\omega^2\delta)
\end{align}
leading to 
\begin{align}
\Delta^{1}_{\varphi}\left[\int_0^{\lambda_R} \fv{K}^2\, d\lambda\right] &= \mathcal{O}(\omega^2\delta)
\end{align}
\end{subequations}
For the second, 
\begin{subequations}
\begin{align}
f &= R(\pphi,\plambda,\pphi,\plambda)\\
f^{(0)} &\sim \mathcal{O}(L^{-2})\\
\frac{\partial f^{(0)}}{\partial x^\mu}&\sim\mathcal{O}(L^{-3})\\
f^{(1)} &\sim \mathcal{O}(\omega^2)
\end{align}
leading to 
\begin{align}
\Delta^{1}_{\varphi}\left[\int_0^{\lambda_R} R(\pphi,\plambda,\pphi,\plambda)\, d\lambda\right] &= \mathcal{O}(\omega^2)
\end{align}
\end{subequations}

% \bibliography{./BibFiles/phyjabb,%
% ./BibFiles/local,%
% ./BibFiles/article,%
% ./BibFiles/phdthesis,%
% ./BibFiles/book,./BibFiles/booklet,./BibFiles/proceedings,%
% ./BibFiles/techreport,./BibFiles/misc,./BibFiles/unpublished}

\end{document}